\documentclass[%
 reprint,
 amsmath,amssymb,
 aps,
]{revtex4-1}

\usepackage{graphicx}
\usepackage{dcolumn}
\usepackage{bm}

\usepackage{cancel}
\usepackage{tikz-feynman}
\tikzfeynmanset{compat=1.1.0}
\allowdisplaybreaks
\begin{document}

\title{Constraints on electroweak gauged unparticle model\\ from  the oblique parameters $S$ and $T$}
\author{Hamza Taibi}
\affiliation{Lab. Phys. Mathématique et Subatomique, Université Mentouri, Constantine 1. Rue de Ein Elbey, Constantine 25000, Algérie}
\author{Nourredine Mebarki}%
\affiliation{Lab. Phys. Mathématique et Subatomique, Université Mentouri, Constantine 1. Rue de Ein Elbey, Constantine 25000, Algérie}

\date{\today}

\begin{abstract}
The oblique parameters $S$ and $T$ are calculated in a gauged unparticles model based on the electroweak group $SU(2)*U(1)$ and it's parameters space is constrained using electroweak precision measurements.

\end{abstract}

\maketitle


\section{\label{sec:level1}Introduction}

The standard model(SM) has been so far in excellent agreement with experiment.
 However, it fails to explain neutrino oscillations,
  dark matter and the origin of baryon asymmetry in the universe.
   Moreover, the hierarchy problem indicates that 
   the SM in its minimal version  cannot describe physics above the weak scale. 
   These inconsistencies and shortcoming of the SM 
   prompted the study of physics beyond the standard model(BSM). A particularly interesting model of BSM proposed about a decade ago is unparticle model \cite{1} wish describe a 
   scale invariant hidden sector interacting with SM particles
    at high energy via messenger particles. These interactions 
    are organized in an effective field theory in wish unparticle are 
    represented by scale invariant operators. An extension of the unparticle model to include operator 
    with quantum number was introduced in \cite{cacciapaglia2008colored}. For any new physics model to be valid it must be consistent with the SM predictions. In this regard the electroweak 
    precision tests represent a powerful tool to test the compatibility of new model with experimental data. To achieve this goal for the unparticle model we  consider unparticle fields embodied in the SM electroweak group. 
    These fields would induce loop effects on the electroweak precision tests represented as contributions to the oblique parameters $S$ and $T$ \cite{peskin1992estimation}.

In section 2 we give a short review of gauged unparticle model and we calculate its contributions to the oblique parameters $S$ and $T$. In section 3 we use the results of the previews section to study the parameters space of unparticles and finally a short summary and conclusion are given.

\section{The model}

 The purpose of our paper is to calculate the effects of unparticles sector on electroweak observables. For this reason we must find Feynman vertices describing the interactions of unparticle fields with the electroweak SM gauge bosons.
 
 The unparticle stuff are described by scale invariant fields with scaling dimension $d$. Conformal invariance impose a particular form for the green function of unparticles. The free propagator of fermionic unparticles in momentum space is:
 \begin{equation}
 \Delta_{U_f}(p,\mu)=\dfrac{A(d-1/2)}{2cos(\pi d)} \dfrac{i}{(\cancel{p}-m) \Sigma_0(p)}
 \end{equation}
where $\Sigma_0(p)=(m^2-p^2 )^{3/2-d}$, $p$ is the momentum, $m$ is the conformal symmetry breaking scale, and $A$ is a normalization factor defined by:
 \begin{equation}
 A(d)=\dfrac{16\pi^{3/2}}{(2\pi)^2d)} \dfrac{\Gamma(d+1/2)}{\Gamma(d-1) \Gamma(2d)}
 \end{equation}
 with $3/2 \leq d \leq 5/2$.
 In order to incorporate the unparicle fields to the SM gauge group we use the following action:
\begin{eqnarray}
 S=\int d^{4}xd^{4}y \left( \mathcal{U}_{L}^{\dagger}(x)\tilde{\Delta}_{\mathcal{U}}^{-1}(x-y)\mathcal{W}_{L}(x,y)\mathcal{U}_{L}\right.\nonumber\\
 +\left.\mathcal{U}_{R}^{\dagger}(x)\tilde{\Delta}_{\mathcal{U}}^{-1}(x-y)\mathcal{W}_{R}(x,y)\mathcal{U}_{R}\right) \label{undoub}
 \end{eqnarray}
 where $\mathcal{U}_{L}$ is unparticle multiplet wish transform according to the gauge group $SU(2)_{L}$. $\mathcal{U}_{R}$ is $SU(2)_{L}$ singlet wish transform according to the hypercharge group $U(1)_{Y}$. To ensure gauge invariance we have introduced the Wilson line $\mathcal{W}(x,y)$ defined as:
  \begin{equation}
  \mathcal{W}_{L}(x,y)=P\exp(\int_{x}^{y}(T^{a}W_{\mu}^{a}-ig^{'}Y B_{\mu})du^{\mu}) \label{Wline1}
  \end{equation}
  \begin{equation}
\mathcal{W}_{R}(x,y)=\exp(\int_{x}^{y}-ig^{'}Q B_{\mu}du^{\mu})\label{Wline2}
  \end{equation}
$P$ denote path ordering in the generators $T^{a}$ in the unparticle representation. $Q$ is the charge operator in the same representation. 
 To find the interaction vertices of unparticles with physical gauge bosons $Z$, $W$ and $\gamma$ of the SM  we replace $W^{a}$, $B$ in Eqs. (\ref{Wline1},\ref{Wline2}) according to the relations:
  \begin{equation}
    W_{3}^{\mu}=\cos(\theta_{W})Z^{\mu}+\sin(\theta_{W})A^{\mu}
  \end{equation}
 
  \begin{equation}
  B^{\mu}=-\sin(\theta_{W})Z^{\mu}+\sin(\theta_{W})A^{\mu}
    \end{equation}
  
  \begin{equation}
 W^{\mu}=(W_{1}+iW_{2})/\sqrt{2},W^{\mu\dagger}=(W_{1}-iW_{2})/\sqrt{2}
  \end{equation}
 where $\theta_{W}$ is the Weinberg mixing angle.
  
  Now using the same techniques developed by Terning et al, in the context of nonlocal chiral quark model(see Ref. \cite{terning1991gauging}), we derive Feynman vertices for the coupling of unparticle with one and two gauge bosons as follows
\begin{align}
\Gamma^{\mu}(p,q)=&i g\bigl( \gamma^{\mu}(T_{a}+T_{b}\gamma_{5})(\Sigma_0(p)+\Sigma_0(p+q))\nonumber\\
&+(2\cancel{p}+\cancel{q}-2m)(T_{a}+T_{b}\gamma_{5})(2p+q)^{\mu}\nonumber\\
&\times\Sigma_1(p,q)\bigr)\label{Fvertex1} 
\end{align}
and
\begin{align}
&\Gamma^{ab\mu\nu}(p,q_{1},q_{2})=i\dfrac{g_{a}g_{b}}{2}\bigl((2\cancel{p}+\cancel{q_{1}}+\cancel{q_{2}})\bigl[ \left\lbrace T^{a},T^{b}\right\rbrace \nonumber\\
&\times g^{\mu\nu}\Sigma_{1}(p,q_{1}+q_{2})+T^{a}T^{b}(2p^{\mu}+2q_{2}^{\mu}+q_{1}^{\mu})\nonumber\\
&\times(2p^{\nu}+q_{2}^{\nu})\Sigma_{2}(p,q_{2},q_{1})+T^{b}T^{a}(2p^{\mu}+q_{1}^{\mu})\nonumber\\
&\times(2p^{\mu}+2q_{1}^{\mu}+q_{2}^{\mu})\Sigma_{2}(p,q_{1},q_{2})\bigr]+\gamma^{\mu}\Gamma^{ab\nu}(p,q_{2},q_{1})\nonumber\\
&+\gamma^{\nu}\Gamma^{ab\mu}(p,q_{1},q_{2})\bigr)
\label{Fvertex2}
\end{align}
$g$ and $g_{a,b}$ denote unparticle coupling  with SM gauge bosons, $T_{a}$ and $T_{b}$ are operators defined in the unparticles representation and the form factors are
  \begin{equation}
\Sigma_{1}(p,q)=\dfrac{\Sigma_{0}(p+q)-\Sigma_{0}(p)}{(p+q)^{2}-p^{2}}\label{Ffactor1}
  \end{equation},
  \begin{equation}
\Sigma_{2}(p,q_{1},q_{2})=\dfrac{\Sigma_{1}(p,q_{1}+q_{2})-\Sigma_{1}(p,q_{1})}{(p+q_{1}+q_{2})^{2}-(p+q_{1}^{2})}\label{Ffactor2}
  \end{equation}
and $\Gamma^{ab\mu}$ is defined as 
  \begin{eqnarray}
\Gamma^{ab\mu}=\lefteqn{T^{b}T^{a}\left(2p^{\mu}+q_{1}^{\mu}\right)\Sigma_{1}(p,q_{2})}\nonumber\\
&+T^{b}T^{a}\left(2p^{\mu}+2q_{2}+q_{1}^{\mu}\right)\Sigma_{1}(p+q_{2},q_{1})
  \end{eqnarray}
  
For the abelin group $U(1)$ it is sufficient to replace $T_{a}$ with 1 and $T_{b}$ with 0. For $W$ and $Z$ we define $T_{a}$ and $T_{b}$ as follows 
  \begin{equation}
    \text{for $W$}: T_{a}=\dfrac{\sigma^{-}}{2},T_{b}=\dfrac{\sigma^{+}}{2}
  \end{equation}
    \begin{equation}
\text{for $Z$}: T_{a}=\dfrac{\sigma^{3}}{2}-2\sin^{2}(\theta)Q,T_{b}=-\dfrac{\sigma^{3}}{2}
  \end{equation}
$\sigma^{-}$,$\sigma^{+}$ and $\sigma^{3}$ are Pauli matrices.

Now that we have derived Feynman vertices we can calculate the unparticle contribution to the oblique parameters $S$ and $T$. The  explicit expressions of these parameters are the following 
    \begin{eqnarray}
S=&&\dfrac{4s_{w}^{2}c_{w}^{2}}{\alpha}\left(\dfrac{\Pi_{ZZ}(m_{Z}^{2})-\Pi_{ZZ}(0)}{m_{Z}^{2}}-\dfrac{c_{w}^{2}-s_{w}^{2}}{s_{w}c_{w}}\right.\nonumber\\
&&\left.\times\Pi_{Z\gamma}^{'}(0) -\Pi_{\gamma\gamma}^{'}(0)\right)\label{Sdef} 
    \end{eqnarray}  
and
 \begin{equation}
 T=\dfrac{1}{\alpha}\left[\dfrac{\Pi_{WW}(0)}{m_{W}^{2}}-\dfrac{\Pi_{ZZ}(0)}{m_{Z}^{2}}\right]\label{Tdef}  
  \end{equation}
  
$\Pi_{ab}(q^{2})$, with $a$,$b$ stand for $\gamma$, $Z$ or $W$, denote the new physics contribution to the  part proportional to the metric $g_{\mu\nu}$ of the  self-energies functions\\
$\Pi_{ab}^{\mu\nu}(q^2)=ig_{\mu\nu}\Pi_{ab}(q^{2})+\dots$. The derivatives $\Pi_{ab}^{'}(q^{2})$  are defined by $\Pi_{ab}^{'}(q^{2})=d\Pi_{ab}(q^{2})/dq^2$. $\alpha$ is the fine structure constant and $s_w=\sin(\theta_W)$, $c_w=\cos(\theta_W)$.
 
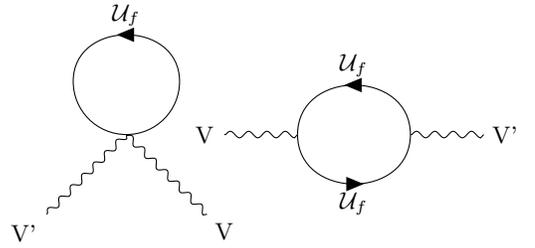
\begin{figure}[!htb]
    \begin{tikzpicture} 
\begin{feynman}
\vertex(a1);
\vertex[above right=of a1] (b1);
\vertex[below right=of b1] (c1){\($V$\)};
\vertex[above right=1cm of b1](d1);
\vertex[above left=1cm of b1 ](e);
\vertex[right=4em of b1](a2);
\vertex[right=3em of a2] (b2);
\vertex[right=of b2](c2) ;
\vertex[right=3em of c2](d2){\($V'$\)} ;

\diagram* {
(a1) -- [photon] (b1)-- [plain,quarter right](d1),
(b1) -- [photon] (c1),(d1) -- [fermion,half right,edge label'=\(\mathcal{U}_f\)] (e),(e) -- [plain,quarter right] (b1),
(a2) -- [photon] (b2)-- [fermion,half right,edge label'=\(\mathcal{U}_f\)](c2),
(c2) -- [fermion,half right,edge label'=\(\mathcal{U}_f\)] (b2),
(c2) -- [photon] (d2),
};
\vertex [left=0.07em of a2] {\($V$\)};
\vertex [below left=0.1em of a1] {\($V'$\)};
\end{feynman} 
\end{tikzpicture}
   \caption{\label{loop}The one loop contribution to polarisation functions from charged fermionic unparticle fields, $V$ and $V'$ stand for $\gamma$, $Z$ or $W$.}
 \end{figure}

 In Fig.\ref{loop} we show a typical diagram of the fermionic unparticle loops contributions to selfenergie functions $\Pi_{ab}(q^{2})$ at the one loop level, where $V$ and $V'$ stand for $\gamma$, $Z$ or $W$ . The complicated expressions that define the Feynman vertices Eqs(\ref{Fvertex1},\ref{Fvertex2}) does not allow the application of Passarino Veltman method to reduce tensor integrals to  simpler scalar integrals. Howover, if we look at the large $p$ region of the loop integral, as is done in \cite{2,3}, we can affect a taylor expansion of the function $\Sigma(p+q)$ for small q. To first order in the expansion coefficient $y=q+2p.q$  the form factors $\Sigma_1$ and $\Sigma_2$,  defined in Eqs. (\ref{Ffactor1},\ref{Ffactor2}), become
    \begin{equation}
\Sigma_1(p,q)\simeq\dfrac{(-1)^{3/2-d}(3/2-d)}{\left( (p+q)^2-m^2)\right) ^{d-1/2}}+\cdots\label{approx1}
    \end{equation}
and 
 \begin{multline}
    \Sigma_2(p,q_1,q_2)\simeq \dfrac{(-1)^{5/2-d}(3/2-d)(d-1/2)}{\left((p+q_1)^2-m^2\right)}\\
\times\dfrac{1}{\left((p+q_1+q_2)^2-m^2\right)^{d-1/2}}+\cdots\label{approx2}
            \end{multline}

Using Eq. (\ref{approx1}) and Eq. (\ref{approx2}) in the calculations of  the loop integrals contained in the polarisation functions of Eq. (\ref{Sdef}) and Eq. (\ref{Tdef}) we find The one loop contribution of unfermions to the oblique parameters $S$ and $T$ as follows
\begin{widetext}
\begin{eqnarray}
T=&&\dfrac{m^2}{8\pi s_{w}^2c_{w}^2M_{Z}^2}\left\lbrace \frac{23}{6}+4d-2d^2+\left( 1+2s_{w}^2(q_d-q_u)+4s_{w}^4(q_{u}^2+q_{d}^2)\right)\left(-\frac{7}{2}+6d-\frac{10}{3}d^2-\frac{2}{3}\left( \frac{3}{2}-d\right)^2\right.\right.\nonumber\\
&&\left.\times\dfrac{1+3s_w^2(q_d-q_u)+6s_w^4(q_u^2+q_d^2)}{1+2s_w^2(q_d-q_u)+4s_w^4(q_u^2+q_d^2)}\right) +\ln\left(\frac{\mu^2}{m^2}\right)\left(-\frac{5}{4}+\frac{14}{2}d-4d^2+\left( 1+2s_w^2(q_d-q_u)+4s_w^4(q_u^2+q_d^2)\right)\right.\nonumber\\
&&\times \left.\left. \left(43-44d+12d^2+\dfrac{4s_{w}^2\left((q_d-q_u)+2s_{w}^2(q_{u}^2+q_{d}^2)\right)}{1+2s_{w}^2(q_d-q_u)+4s_{w}^4(q_{u}^2+q_{d}^2}\right)\right)\right\rbrace
\end{eqnarray}
and
\begin{eqnarray}
S=&&-\dfrac{\left( 1+2s_{w}^2(q_d-q_u)+4s_{w}^4(q_{u}^2+q_{d}^2)\right)}{2\pi}\Biggl( F_1+F_2+\ln \left(\frac{\mu^2}{m^2}\right)(F_3+F_4+F_5)\Biggr)
+\dfrac{c_{w}^2-s_{w}^2}{48\pi}\nonumber\\
&&\times\Biggl(\dfrac{q_d-q_u+2s_{w}^2(q_{u}^2+q_{d}^2)}{70}\left(6115-6219d+568d^2+420d^3\right)+8(q_d-q_u)(3/2-d)^2(d-1/2)+\ln\left(\frac{\mu^2}{m^2}\right)\nonumber\\
&&\times\left(q_d-q_u-4s_{w}^2(q_{u}^2+q_{d}^2)\right)\left(-16-33d+28d^2-4d^3\right)\Biggr) +\dfrac{s_{w}^2c_{w}^2(q_{u}^2+q_{d}^2)}{24\pi} \Biggl(\dfrac{7123-4959d+848d^2-140d^3}{70}\nonumber\\
&&+\ln\left(\frac{\mu^2}{m^2}\right)\left(-16-33d+28d^2-4d^3\right)\Biggr) 
\end{eqnarray}
with 
\begin{eqnarray}
F_1=&&\frac{1}{3}\left( d-\frac{1}{2}\right)^2\left( \frac{5}{2}-d \right){}_{4}{F}_{3}\left( 1,1,d+\frac{1}{2},\frac{7}{2} , 2,2,\frac{5}{2} ,\frac{\tau}{4}\right)+\frac{1}{4}\left( d+\frac{1}{2}\right)\left( \frac{5}{2}-d \right)\left( \frac{3}{2}-d \right)\nonumber\\
&&\times{}_{4}{F}_{3}\left( 1,1,d+\frac{3}{2},\frac{7}{2}-d , 2,3,\frac{5}{2} ,\frac{\tau}{4}\right)+\left( \frac{3}{2}-d \right)^2\left( \left( \frac{4}{3}\frac{m^2}{M_{Z}^2}-\frac{11}{24}\right){}_{2}{F}_{1}\left( 1,2,\frac{5}{2},\frac{\tau}{4}\right)-\frac{4}{3}\frac{m^2}{M_{Z}^2}\right.\nonumber\\
&&+\left.\frac{1}{5}\left({}_{2}{F}_{1}\left( 1,3,\frac{7}{2},\frac{\tau}{4}\right)+8{}_{3}{F}_{2}\left( 1,1,3 , 2,\frac{7}{2}, \frac{\tau}{4}\right)\right)\right)
\end{eqnarray}
\begin{eqnarray}
F_2=&&-\left(\frac{3}{2}-d\right)^2\left(\left( \frac{2}{3}\frac{m^2}{M_{Z}^2}-\frac{1}{3}\right)\dfrac{1+3s_{w}^2(q_d-q_u)+6s_{w}^4(q_{u}^2+q_{d}^2)}{1+2s_{w}^2(q_d-q_u)+4s_{w}^4(q_{u}^2+q_{d}^2)}{}_{2}{F}_{1}\left( 2,2,\frac{5}{2},\frac{\tau}{4}\right)+\frac{1}{30}(8-\tau)\right.\nonumber\\
&&\times{}_{2}{F}_{1}\left( 2,3,\frac{7}{2},\frac{\tau}{4}\right)\left. +\frac{1}{35}{}_{2}{F}_{1}\left( 2,4,\frac{9}{2},\frac{\tau}{4}\right)-\dfrac{2m^2}{3M_{Z}^2}\dfrac{ 1+3s_{w}^2(q_d-q_u)+6s_{w}^4(q_{u}^2+q_{d}^2)}{1+2s_{w}^2(q_d-q_u)+4s_{w}^4(q_{u}^2+q_{d}^2)}\right)\nonumber\\
 &&+ \left(\frac{1}{12}+\frac{13}{6}d-\frac{4}{3}d^2\right){}_{2}{F}_{1}\left( 1,1,\frac{5}{2},\frac{\tau}{4}\right)
\end{eqnarray}
\begin{eqnarray}
F_3=&&\frac{1}{6}\left( \left( d-\frac{1}{2}\right)\left( \frac{5}{2}-d\right){}_{3}{F}_{2}\left( 1,d+\frac{1}{2},\frac{7}{2}-d , 2,\frac{5}{2} ,\frac{\tau}{4}\right)+{}_{2}{F}_{1}\left( 1,2,\frac{5}{2},\frac{\tau}{4}\right)\right)+\left(\frac{3}{2}-d\right)^2\nonumber\\
&&\times\left(\left( -4\frac{m^2}{M_{Z}^2}+\frac{\tau}{4}\right){}_{2}{F}_{1}\left( 1,2,\frac{5}{2},\frac{\tau}{4}\right)-\frac{6}{5}{}_{2}{F}_{1}\left( 1,3,\frac{7}{2},\frac{\tau}{4}\right)+4\frac{m^2}{M_{Z}^2}\right)
\end{eqnarray}
\begin{eqnarray}
F_4=&&\left(\frac{3}{2}-d\right)\left(\left( \frac{m^2}{M_{Z}^2}-\frac{1}{2}\right) \left({}_{3}{F}_{2}\left( 1,d+\frac{1}{2},\frac{5}{2}-d,2,\frac{3}{2},\frac{\tau}{4}\right)+{}_{2}{F}_{1}\left( 1,1,\frac{3}{2},\frac{\tau}{4}\right)\right)-2\frac{m^2}{M_{Z}^2}\right.\nonumber\\
&&+\left.\frac{2}{\Gamma(5)}\left(\left(d+\frac{1}{2}\right)\left(\frac{5}{2}-d\right){}_{3}{F}_{2}\left(1,d+\frac{3}{2},\frac{7}{2}-d,3,\frac{5}{2} ,\frac{\tau}{4}\right)+2{}_{2}{F}_{1}\left( 1,2,\frac{5}{2},\frac{\tau}{4}\right)\right)\right)  
\end{eqnarray}
\begin{eqnarray}
F_5=&&\dfrac{2s_{w}^2\left(q_d-q_u+2s_{w}^2(q_{u}^2+q_{d}^2)\right)}{1+2s_{w}^2(q_d-q_u)+4s_{w}^4(q_{u}^2+q_{d}^2)}\left( {}_{3}{F}_{2}\left(1,d-\frac{1}{2},\frac{5}{2}-d, 1,\frac{3}{2},\frac{\tau}{4}\right)+{}_{2}{F}_{1}\left( 1,1,\frac{3}{2},\frac{\tau}{4}\right)-2\right)
\end{eqnarray}

\end{widetext}
where $\displaystyle{\tau=M_{Z}^2/m^2}$ and ${}_{2}{F}_{1}$, ${}_{3}{F}_{2}$ and ${}_{4}{F}_{3}$ are hypergeomtric functions. $\mu$ is the renormalization scale constant. In general $\mu$ takes arbitrary values but since we are working with experimental data extracted at the LEP experiments, with momentum scale around the Z pole, we choose in the following study values of  $\mu$ in the range  $ \mu\in\left[ M_{Z}/2,2M_{Z}\right]$.

\section{phenomenology}
\begin{figure}[!htb]
 \includegraphics[scale=1]{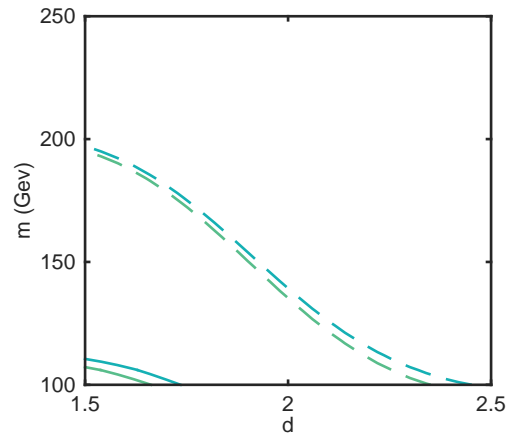}
  \caption{\label{Tcountor} countour plots in the plane (d,m) for $T=0,13$ represented by the upper solid and dashed lines and $T=-0,13$ represented by the lower solid and dashed lines,solid lines are countour plots for $\mu=M_Z$ and dashed lines are for $\mu=2M_Z$.}
  \end{figure}

In order to find the region of parameter space of unparticles that is compatible with experimental limits we must compare the unparticle contributions to the oblique  parameters $S$ and $T$ to the fitted values deduced by comparing the theoretical predictions  of the electroweak observables in the SM and their experimental values \cite{baak2013global}. The fitted values of $S$ and $T$ are the following 
\begin{eqnarray}
\Delta S&=&S-S_{SM}=0,05\pm 0,11\nonumber\\
\Delta T&=&T-T_{SM}=0,09\pm 0,13\label{expbound}
\end{eqnarray}
To illustrate the bounds on unparticle parameters from electroweak precision tests we present in Fig. \ref{Scountor1} and Fig. \ref{Tcountor} contour plots in the  plane of $(d,m)$ in the regions $1,5\leq d \leq 2,5$ and $100\leq m \leq 1100$ for $S$ and $100\leq m \leq 250$ for $T$. In this study we have chosen the values $q_u=-1$, $q_d=0$ for the charges of the upper and lower components, respectively, of the unparticle multiplet introduced in Eq. (\ref{undoub}). In Fig. \ref{Scountor1} countour plots for experimental upper and lower bounds  $S=0,11$ and $S=-0,11$ are depicted for two choices of the renormalisation scale $\mu$. For $\mu=M_Z/2$ the solid line in the right hand side represent $S=0,11$  and the solid line in the left hand side represent $S=-0,11$. For $\mu=2M_Z$ the dashed line in the right represent $S=0,11$  and the dashed line in the left represent $S=-0,11$. As can bee seen from this figure, for values of scale dimension $d\leq 1,7$ there is practically no constraints on the values of conformal breaking scale $m$ but for values of $d\geq 1,7$ $m$ is restricted to values $m\leq 200 Gev$. The allowed region in the parameters space $(d,m)$ become narrower as $d$ increases. For $\mu=2M_Z$ the scale dimension $d$ must be inferior to 1,7 to satisfies the experimental bounds.
\begin{figure}[!htb]
 \includegraphics[scale=1]{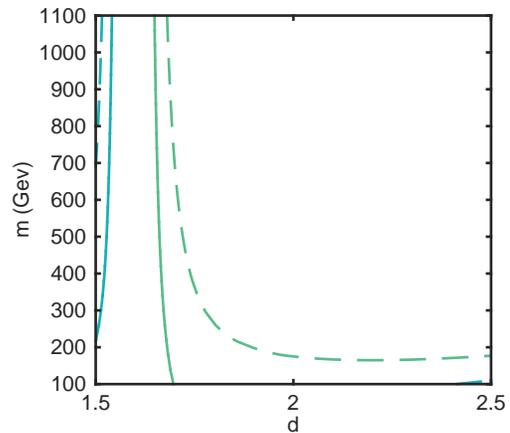}
  \caption{\label{}countour plots in the plane (d,m) for $S=0,11$ on the right hand side and $S=-0,11$ on the left hand side,solid lines are countour plots for $\mu=2M_Z$ and dashed lines are countour plots for $\mu=M_Z/2$.}
  \end{figure}
Fig. \ref{Tcountor} shows countour plots for the upper and lower experimental limits $T=0,13$(the upper lines )  and $T=-0,13$(the lower lines). The solid plots represent $T$ for the renormalisation scale value $\mu=100 Gev$ and the dashed plots represent $T$ for $\mu=2M_Z$. The region between the two solid  lines and the two dashed lines are consistent with measurements for the choosing renormalisations scale value. It is clear from this figure that the oblique parameter $T$ imposes a strong constraint on the allowed region of parameters space. For $\mu=2M_Z$
values of the conformal breaking scale $m\geq 200$ are excluded in the range $1,5\leq d \leq 2,5$. For $\mu=100 Gev$ the allowed region is smaller. The allowed values of the scale dimension $d$ shrink to the range $1,5\leq d \leq 1,7$ and $m\leq 110$.

\begin{figure}
 \includegraphics[scale=1]{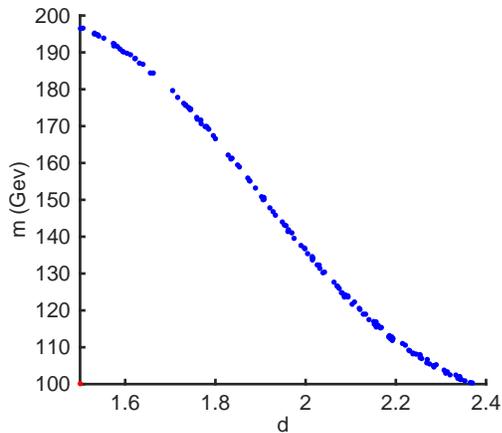}
  \caption{\label{scatterplot}scatter plot in the plane $(d,m)$ wish show the region in parameters space compatible with the $1\sigma$ experimental bound.}
  \end{figure}

Fig. \ref{Scountor1} and Fig. \ref{Tcountor} are based on the bounds expressed by Eq. (\ref{expbound}) in wish $S$ and $T$ are taken as independent parameters. In reality there is a correlation between these two observables expressed by the correlation coefficient $\rho=0,9$ \cite{baak2013global}. Fig. \ref{scatterplot} shows scatter plots in the $(d,m)$ plane compatible with $1\sigma$ experimental bounds  of electroweak precision data in wish  the correlation coefficient $\rho$ is taken into account. The blue dots represent scatter points  for the renormalization scale value $2M_Z$. The red point represent the allowed region for $ \mu=M_Z$. From this figure we see that the allowed region is highly sensitive to the value of the renormalisation scale in the chosen range. The IR cutuf scale $m$ is constrained to values $m\leq200$ but the scale dimension can take value up to 2.34 for $\mu=2M_Z$. In general the combined fitted results of $S$ and $T$, expressed by Fig. \ref{scatterplot}, are compatible with the restrictions imposed by the oblique parameter $T$ (Fig. \ref{Tcountor}) except that the allowed region get smaller in the edges, when $d$ approaches 2,4 and the conformal breaking scale $m$ approaches 200.

\section{Conclusion and Summary}
In this work we have calculated the contribution of a gauged unpaticle model, based on the electroweak group $SU(2)\times U(1)$, to the oblique parameters $S$ and $T$. We have used the results of this calculation to construct the region in the parameters space $(d,m)$ consistent with electroweak precision measurements represented by $S$ and $T$. For different choices of the renormalisation scale constant we have found that the conformal breaking scale must be $m\leq200 Gev$ for $1,5\leq d \leq 2,4$ in order to satisfies the experimental bounds.

\section*{Data availability}
Only analytical investigations (no data) were used to support the findings of this study.

\acknowledgments

This work is supported by the Algerian Ministry of High Education and Scientific Research.

\nocite{*}

\bibliography{refs}

\end{document}